\documentclass[11pt]{article}

\pdfoutput=1
\usepackage{natbib}
\usepackage{verbatim}
\usepackage{fixltx2e}
\usepackage{amsfonts, amsmath, amssymb} 
\usepackage{graphicx} 
\usepackage[colorlinks]{hyperref}
\usepackage{authblk}
\providecommand{\e}[1]{\ensuremath{\times 10^{#1}}}

\title{Contrast Agent Quantification by Using Spatial Information in Dynamic Contrast Enhanced MRI}
\author[1]{Jianfeng Wang\thanks{jianfeng.wang@umu.se}}
\author[2]{Anders Garpebring}
\author[2]{Patrik Brynolfsson}
\author[1]{Xijia Liu}
\author[1]{Jun Yu}
\affil[1]{Department of Mathematics and Mathematical Statistics, Ume\aa { }University, Ume\aa , Sweden}
\affil[2]{Department of Radiation Physics, Ume\aa { }University, Ume\aa , Sweden}

\begin{document}

\maketitle

\begin{abstract}
The purpose of this study is to investigate a method, using simulations, to improve contrast agent quantification in Dynamic Contrast Enhanced MRI. Bayesian hierarchical models (BHMs) are
applied to  smaller images ($10\times10\times10$) such that spatial information can be incorporated. Then exploratory analysis is done for larger images ($64\times64\times64$) by using
maximum a posteriori (MAP).

For smaller images: the estimators of proposed BHMs show improvements in terms of the root mean squared error compared to the estimators in existing method for a noise level equivalent of
    a 12-channel head coil at 3T. Moreover, Leroux model outperforms Besag models. For larger images: MAP estimators also show improvements by assigning Leroux prior.\\

\noindent keywords: Contrast agent quantification, BHM, Besag, Leroux, INLA, MAP
\end{abstract}

\section{Introduction}
With dynamic contrast-enhanced MRI (DCE-MRI) the uptake and washout of a exogenous contrast agent (CA) can be monitored in certain tissues, e.g. tumors. By analysing the dynamics of the CA
concentration using pharmacokinetics models \citep{Sourbron2013}, it is possible to estimate physiology parameters such as blood flow, vessel density, capillary endothelial permeability,
and extravascular extracellular space volume. These parameters are useful for characterizing e.g. tumor angiogenesis \citep{Verma2012}, performing target delineation and evaluating
treatment response in radiotherapy \citep{Cao2011}.

A crucial step for successful parameter estimation is accurate determination of the CA concentration. This is difficult since the MR-signal have a complicated relationship to the CA
concentration through the effect of the CA on the tissue relaxation time constants. Most commonly, CA concentration is estimated using the magnitude of the MRI images
\citep{Sourbron2011}, which is referred to as magnitude estimated CA in this paper. However, the accuracy of this method can be hampered by for instance issues like flip angle inhomogeneity
\citep{Cheng2007}.

In addition to the magnitude information, MRI images also contain phase information. And the phase is influenced by the CA concentration. This has been exploited for accurate blood CA
estimates and more recently Brynolfsson et al.. \citet{Brynolfsson2014} combined magnitude and phase information to improve CA estimation in all types of tissue. In the work by
Brynolfsson et al., a maximum likelihood estimator (MLE) was used to find the most likely CA concentration given noisy and biased magnitude images and noisy phase images. It is likely that
CA concentration estimates can be improved even further if spatial (neighbor) information could be incorporated, since it is natural to assume that voxels next to each other behave alike,
especially for the voxels in the same tissue.

The aim of this work is therefore to explore whether spatial information can provide further improvement by combining it with phase shift data and magnitude estimated CA. At current stage,
we limit our evaluations to simulation data, since it is essential to establish the feasibility before trying to solve more difficult practical issues, such as the background phase drift.

The structure of the paper is organized as follows. The models with and without spatial information will be introduced in Section 2. In Section 3, the methods used in this work will be
specified. Results will be presented in Section 4. The paper will be closed with conclusion and discussion.

\section{Theory}
\subsection{Magnitude and phase models}
This work uses the model developed by Brynolfsson et al. \citep{Brynolfsson2014} as its starting point. Briefly, the model assumes that a measurement of the CA concentration
$\mathbf{c}=(c_1,...c_N)$, where $N$ is the number of voxels in the image, is available from magnitude estimated CA data $\mathbf{c}_m=(c_1,...,c_N)$ and in addition to that also available
from phase shift data $\Delta \mathbf{\varphi}=(\Delta \varphi_1,...,\Delta \varphi_N)$. The statistical model connecting the underlying CA concentration $\mathbf{c}$ and the measured data
has the form
\begin{align*}
\mathbf{c}_m&=D_{\boldsymbol{\xi}}\mathbf{c}+\boldsymbol{\epsilon}_m \tag{1}\\
\Delta\boldsymbol{\varphi}&=\Psi\mathbf{c}+\boldsymbol{\epsilon}_\varphi \tag{2},\\
\end{align*}
where $\boldsymbol{\epsilon}_m$, $\boldsymbol{\epsilon}_{\varphi}$ are two Gaussian white noise vectors and $D_{\boldsymbol{\xi}}$ is a diagonal matrix with diagonal elements
$\boldsymbol{\xi}$ which is a Gaussian white noise vector with mean value $\mathbf 1$. Thus, the magnitude estimated CA is both noisy and biased, while the phase shift data is only noisy.
The matrix $\Psi$ represents a convolution and corresponds to how the magnetic properties of the CA perturb the measured phase shift data. An important feature of $\Psi$ is that it is
singular and thus cannot be inverted. Equivalently, the matrix vector notation $(2)$ can be written voxel wise given by
\begin{align*}
\Delta\varphi(\mathbf{s})=F^{-1}(G(\mathbf{k})\cdot F(c(\mathbf{s})))+ \epsilon_\varphi(\mathbf{s})\tag{2*},
\end{align*}
where $F$ is the Fourier transform function, $\mathbf{s}$ is the position vector in the image, $\mathbf{k}$ is the coordinate position vector in $k$-space and $G$ is a known function of
$\mathbf{k}$ \citep{Brynolfsson2014}. Note that the drawback of $(2)$ compared with $(2^*)$ is that $\Psi$ could be too huge to be stored into computer's memory.
\subsection{Model I: No spatial dependence}
No spatial dependence is assumed to $\mathbf c$ in this model implying no spatial prior is added to the model $(1)$ and $(2)$ for  $\mathbf c$. In this case $\mathbf{c}_m$ and
$\Delta\boldsymbol{\varphi}$ are assumed to be independent between them and the MLE
\begin{align*}
\hat{\mathbf{c}}&=\underset{\mathbf{c}}{\mathrm{argmin}} \left\{(\mathbf{c}_m-\mathbf{c})^T\Sigma_m^{-1}(\mathbf{c}_m-\mathbf{c})+(\Delta\boldsymbol{\varphi}-\Psi\mathbf{c})^T
\Sigma_\varphi^{-1}(\Delta\boldsymbol{\varphi}-\Psi\mathbf{c})\right\}
\end{align*}
is used to find $\mathbf{c}$ \citep{Brynolfsson2014}.

\subsection{Model II: Bayesian hierarchical models (BHMs)}
To incorporate spatial information into the model $(1)$ and $(2)$, BHM is used to model the spatial relationship. BHM is a statistical model consisting of multiple-stages. It estimates the
parameters of interest by using Bayesian method. It is common to have three stages in the model. The first stage is data model used to model observed data, the second stage is process model
used to model the unknown parameters of interest in data model and the last one is hyperparameter model used to model unknown hyperparameters. There are many well developed process models,
e.g. Besag model \citep{Besag1991},
 BYM model \citep{Besag1991}, Cressie model \citep{Stern2000} and Leroux model \citep{Leroux2000}, two of which are used in this work.
\subsubsection{Besag model}
Besag model is the simplest and most popular model, which has a special form of a generalized model \citep{Besag1974} given by
\begin{align*}
\mathbf{c}\sim \mathcal{N}\left(\mathbf{0},\sigma^2(D(I-\rho W))^-\right), \tag{3}
\end{align*}
where $\sigma^2$ is a variance parameter, $I$ is a identity matrix, $D=diag\left(d_1,...,d_N\right)$ is a known diagonal matrix with $d_i$ is the number of neighbors of voxel $i$,
$W=(W_{ij})$ is the proximity matrix,
\begin{align*}
W_{ij}=\begin{cases}
1/d_i, & i\sim j\\
0, &\text{otherwise}
\end{cases}
\end{align*}
 where $i \sim j$ indicates that the two voxels $i$ and $j$ are neighbors, - represents generalized inverse,
 and $\rho$ is a spatial dependence parameter.
 It can be shown that it suffices to let $\rho \in (1/\min_i\lambda_i, 1)$ to ensure the covariance matrix of $(3)$ to be positive definite,
 where $\lambda_i, i=1,...,N,$ are eigenvalues of $W$ \citep{Reber1999}. However, since $\rho=1$ for Besag model, the covariance matrix of $(3)$ exists only in terms of generalized inverse.

In terms of full conditionals, the model $(3)$ can be expressed as
\begin{align*}
c_i|\mathbf{c}_{j|_{i\sim j}}, \sigma, \rho \sim \mathcal{N}(\frac{\rho}{d_i}\underset{i\sim j}{\sum}c_j, \frac{1}{d_i}\sigma^2), \tag{4}
\end{align*}
where $\mathbf{c}_{j|_{i\sim j}}$ represents the elements which are neighbors of $c_i$. The conditional mean is affected by its neighbors, and conditional variance is proportional to the
variance parameter $\sigma^2$. As mentioned above, $\rho=1$, thus there is no proper joint distribution from which $(4)$ can be derived. However, a sum to zero constraint, $\sum c_i=0$, can
be added to $\mathbf{c}$ to guarantee the identifiability of this random field \citep{Assun2009,Rue2005}.
 The inference process to obtain the estimates will be described in 2.3.3.

\subsubsection{Leroux model}
Although being invariant to the addition of any constant is a very important property \citep{Rue2005}, Besag model has some undesired properties, e.g. the covariance matrix is not positive
definite and it leads to a negative pairwise correlation for regions located further apart \citep{MacNab2010}. A proper prior is introduced here which was proposed by Leroux et al.
\citep{Leroux2000} and was the most appealing from both theoretical and practical standpoints \citep{Lee2011}. The joint distribution of Leroux model is given by
\begin{align*}
\mathbf{c}\sim \mathcal{N}\left(\mathbf{0},\sigma^2\left((1-\lambda)I+\lambda R\right)^{-1}\right), \tag{5}
\end{align*}
where R is the structure matrix for Besag model which equals to $D(I-W)$, $\lambda$ is a spatial dependence parameter taking values within the interval of $(0,1)$. As $\lambda \to 1^{-}$,
the model converges to Besag model and as $\lambda \to 0^{+}$, it converges to $\mathcal{N}(0,\sigma^2I)$.

In terms of full conditionals, the model of $(5)$ can be expressed as
\begin{align*}
c_i|\mathbf{c}_{j|_{i\sim j}}, \sigma, \lambda \sim \mathcal{N}(\frac{\lambda}{1-\lambda+\lambda d_i}\underset{i\sim j}{\sum}c_j, \frac{1}{1-\lambda+\lambda d_i}\sigma^2).
\end{align*}

The major difference between the models in model I and model II is that $\mathbf{c}$ is considered as a deterministic unknown vector in model I while as a random unknown vector in model II
by assuming it is a Gaussian Markov random field (GMRF).
\subsubsection{INLA: Integrated nested Laplace approximations}
To estimate the random unknown vector in BHMs, an algorithm based on INLA has been well developed. This method can be used for GMRFs which has been being applied in many scientific fields.
The BHMs, described above, fit into this frame and are built with three stages. For simplicity reason, a new set of notations of random vectors is introduced, which has no connection with
the notations existing before. The first stage is the data model $\pi (\mathbf{y|x}),$ where $\pi$ denotes probability density, $\mathbf y$ is the observation vector, $\mathbf x$ is the
GMRF and $y_i, i=1,...,N$, are independent conditional on $\mathbf x$. The second stage is the GMRF, $\pi (\mathbf{x}|\boldsymbol{\theta})$, where $\boldsymbol{\theta}$ is the
hyperparameter vector and $\pi(\boldsymbol \theta)$ is the third stage. INLA can provide accurate estimations for the GMRF and hyperparameters. The inference process is described briefly as
follows:

The main interest is to estimate the marginal posterior distributions of the GMRF
\begin{align*}
\pi(x_i|\mathbf{y})=\int_{\boldsymbol{\theta}}\pi(x_i|\mathbf{y},\boldsymbol{\theta})\pi(\boldsymbol{\theta}|\mathbf{y}) \mathrm d\boldsymbol\theta, \tag{6}
\end{align*}
of which the integrand can be obtained using approximations as follows.
\begin{align*}
\hat\pi(\boldsymbol{\theta}|\mathbf{y})\propto \frac{\pi(\mathbf{x,y,}\boldsymbol{\theta})}{\hat{\pi_G}(\mathbf{x|y,}\boldsymbol{\theta})}|_{\mathbf{x=x^*(\boldsymbol{\theta})}},
\end{align*}
where the denominator $\hat{\pi_G}(\mathbf{x|y,}\boldsymbol{\theta})$ denotes the Gaussian approximation to the full conditional distribution of $\mathbf x$, and $x^*(\boldsymbol{\theta})$
is the mode of the full conditional of $\mathbf{x}$ for a given $\boldsymbol{\theta}$. Gaussian approximation means the distribution of a variable is approximated by a normal distribution
by matching the mode and the curvature at the mode \citep{Rue2005}.

The simplified Laplace approximation method is used to approximate the other component of the integrand of $(6)$ \citep{Rue2009}. This method is a trade off between accuracy and
computational time and is commonly used in practice. It is also the default method in R-INLA. In order to perform a numerical integration of $(6)$, a number of good evaluation points
$\boldsymbol{\theta}_k$ of $\boldsymbol \theta$ can be obtained by Newton like algorithms \citep{Rue2009}. Finally, an approximation of the posterior marginal density $(6)$ is given by
\begin{align*}
\pi(x_i|\mathbf{y})=\sum_k \hat{\pi}(x_i|\mathbf{y},\boldsymbol{\theta}_k)\hat{\pi}(\boldsymbol{\theta}_k|\mathbf{y}) \Delta\boldsymbol{\theta}_k
\end{align*}

\subsection{Maximum a posteriori (MAP)}
Due to the limitation of R-INLA in this work describing in the next section, an exploratory analysis is done by using MAP estimator given by
\begin{align*}
\hat{\mathbf{c}}&=\underset{\mathbf{c}}{\mathrm{argmin}}\left\{(\mathbf{c}_m-\mathbf{c})^T\Sigma_m^{-1}(\mathbf{c}_m-\mathbf{c})+
(\Delta\boldsymbol{\varphi}-\Psi\mathbf{c})^T
\Sigma_\varphi^{-1}(\Delta\boldsymbol{\varphi}-\Psi\mathbf{c})+\mathbf{c}^TQ\mathbf{c}\right\} \tag{7},
\end{align*}
where $Q$ is the precision matrix of $\mathbf{c}$. Linear conjugate gradient algorithm is applied to find $\mathbf{c}$.
\section{Methods}
\subsection{Data preparation}
In this simulation study, the data was produced by using simulated GRE based DCE-MRI scans at 3T with a noise level equivalent of a 12-channel head coil (rSNR=5) and with a noise level of a
2-channel body coil (rSNR=1), respectively. rSNR is defined as rSNR$=\eta \cdot$SNR\textsubscript{bodycoil} and $\eta \in[1,5]$. See \citealt{Brynolfsson2014} for details.

R-INLA is used to implement BHMs which has been mentioned in Section 2.3, however, it does not adapt for Fourier transform function used in $(2^*)$, thus BHMs have to be applied to smaller
images ($10\times10\times10$) such that $\Psi$ in $(2)$ can be stored into the computer's memory. Then exploratory analysis is done for larger images ($64\times64\times64$) by utilising the
estimates from smaller images. 30 simulations for smaller images and 50 simulations for larger images were produced and estimates of CA concentration were obtained with 2-second temporal
resolution for the first 30 seconds and 5-second temporal resolution for the last 30 seconds. In other words, there are 22 time points for each simulation.
\subsection{Common settings for all the models}
Time is assumed to be independent. The covariance matrixes of $\boldsymbol{\epsilon}_{m}$ and $\boldsymbol{\epsilon}_{\varphi}$ are approximated from the simulated data, $\boldsymbol\xi$ is
assumed to be $\mathcal{N}(1,0.09I)$ \citetext{see \citealt{Brynolfsson2014} for details}.
\subsection{Specific settings for BHMs}
$(1)$ and $(2)$ are used as data model. The prior for $\log(1/\sigma^2)$ is set to be $\text{Log-Gamma}\left(1,5\e{-5}\right)$ for Besag and Leroux models, which gives higher probability to
relatively smaller variance. The prior for $\mathrm{logit}\lambda$ is set to be $\text{Logitbeta}(1,1)$ for Leroux model, which represents a non-informative prior.

The first order neighbourhood is used in proximity matrix. Two different assumptions are made to construct the proximity matrix $W$ for Besag model. The first is that two voxels next to
each other are not neighbors if they are from different tissues. However, in reality we have no information about tissue classification, thus in the other case we do not give tissue
restriction to $W$ that is two voxels next to each other are always neighbors. Only the second assumption is used for Leroux model. Under each assumption, the precision matrix Q has at most
6 non zero elements in the off-diagonal positions for each row such that Q is a very sparse matrix.

\subsection{From smaller to larger images}
Although we use R-INLA to analyze smaller images, larger images is used in clinic. MAP estimator is used for larger images with Leroux model for $\mathbf{c}$, which implies
$Q^{-1}=\hat\sigma^2((1-\hat\lambda)I+\hat\lambda R)^{-1}$ in $(7)$. By assuming that the spatial dependence $\lambda$ is invariant over different image sizes, the same estimates
$\hat\lambda$'s for smaller images over the time points can be used. Note that full conditional variance $\text{Var}(c_i|\mathbf{c}_{-i})$ is proportional to $\hat\sigma^2$. Since the
resolution of smaller images is lower than that of larger images, the full conditional variance of smaller images should smaller than that of larger images. Therefore, $\hat\sigma^2$ should
be larger than the one in smaller images for each time point in general. Since the average precision $\hat\tau=1/\hat\sigma^2$ over 30 simulations for each time point could be calculated
for smaller images, one can go through all possible values which are smaller than $\hat\tau$ to minimize $(7)$ for larger images. However, for the exploratory purpose, we set $\hat\tau$ to
be a constant, which is smaller than the smallest averaged $\hat\tau$ over all the time points for smaller images, for larger images over all the time points.

To verify whether the spatial information can improve the estimation for lower rSNR, the same procedure is also applied to data produced with rSNR=1 at 3T. Since full conditional variance
of the images with rSNR=1 should be larger than the one with rSNR=5, $\hat\tau$ should be smaller than the one with rSNR=5 for each time point in general. Again, a constant $\hat\tau$ for
images with rSNR=1 is set for all the time points.
\section{Results}
\subsection{Smaller images}

\begin{figure}
    \centering
    \includegraphics[width=0.8\linewidth]{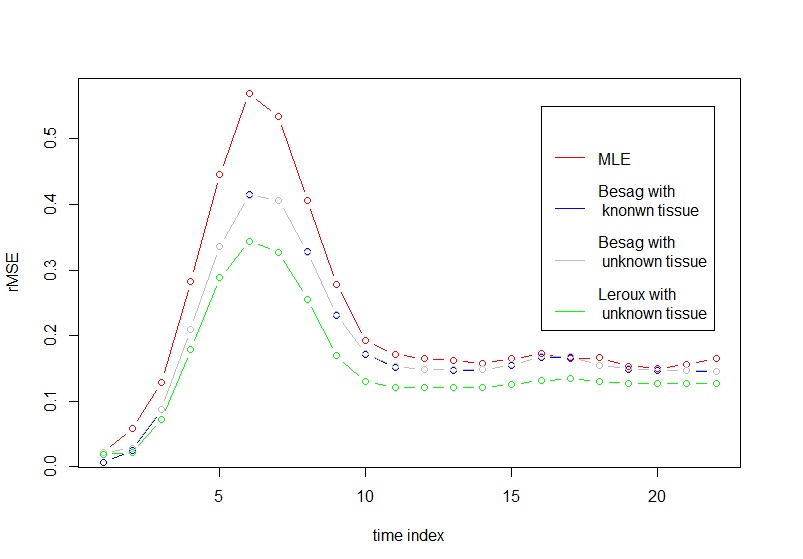}
    \caption{\small {rMSE of vessels at 3T with rSNR=5}}
    \label{fig:2}
\end{figure}
Figure 1 shows root mean squared error (rMSE) of vessels based on 30 simulations. From the figure, two Besag models show improvements in terms of rMSE, especially around the peak, where the
percentage decrease is about 27\%. And the Besag without tissue restriction is slightly worse than the one with tissue restriction, especially at the beginning of the time point. The mean
difference between two Besag rMSE over 22 time points is about 0.001. The Leroux model outperforms the others. The improvements show both around the peak and right tail. It is about 40\%
decrease at the peak and 23\% around the right tail compared to the MLE in model I.

\begin{figure}
    \centering
    \includegraphics[width=0.8\linewidth]{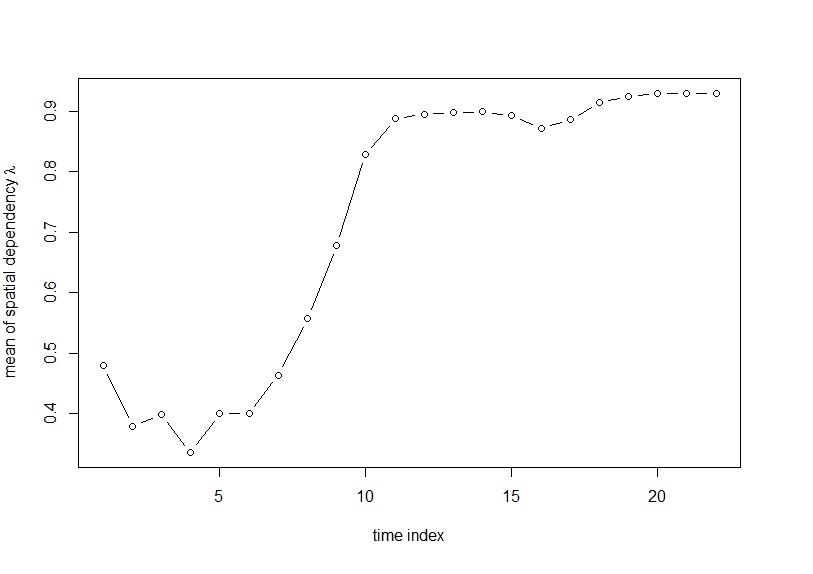}
    \caption{\small {Time series of the mean of spatial dependence $\lambda$} at 3T with rSNR=5}
    \label{fig:3}
\end{figure}

From Figure 2, it shows apparently that the spatial dependence is negatively associated with CA concentration. The spatial dependence is lower when CA concentration is around the peak and
the spatial dependence is getting higher when CA concentration is lower.

\subsection{Larger images}
\begin{figure}
    \centering
    \includegraphics[width=0.8\linewidth]{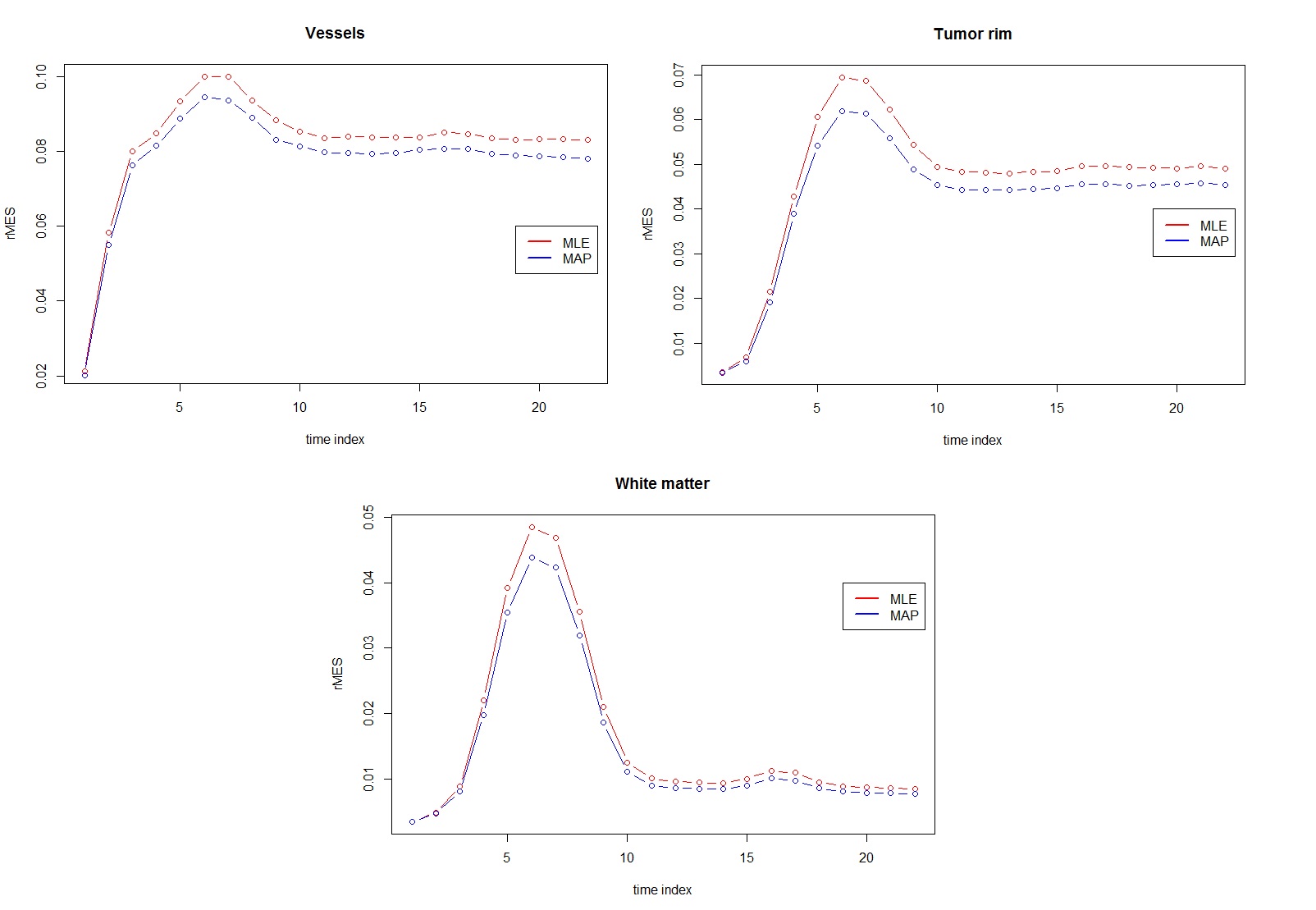}
    \caption{\small {rMSE comparison between MAP estimators and MLEs for vessels, tumor rim and white matter at 3T with rSNR=5 over 50 simulations}}
    \label{fig:4}
\end{figure}
Since the smallest averaged $\hat\tau$ for smaller images over all the time points is 0.9, $\hat\tau=0.1$ is selected to minimize $(7)$ for larger images. The same values shown in Figure 2
are used for $\hat\lambda$ to minimize $(7)$. From Figure 3, it shows both improvements around the peak and right tail for MAP estimators compared to MLEs over 50 simulations, even though
$\hat\sigma^2$s are not the optimal ones to minimize $\mathbf{c}$ over time points.

The same analysis is done for the data produced with rSNR=1 at 3T. It shows that $\lambda$ has similar patten as in Figure 2. $\hat\tau=0.01$, which is less than 0.1, is selected for rSNR=1
at 3T.
\begin{figure}
    \centering
    \includegraphics[width=0.8\linewidth]{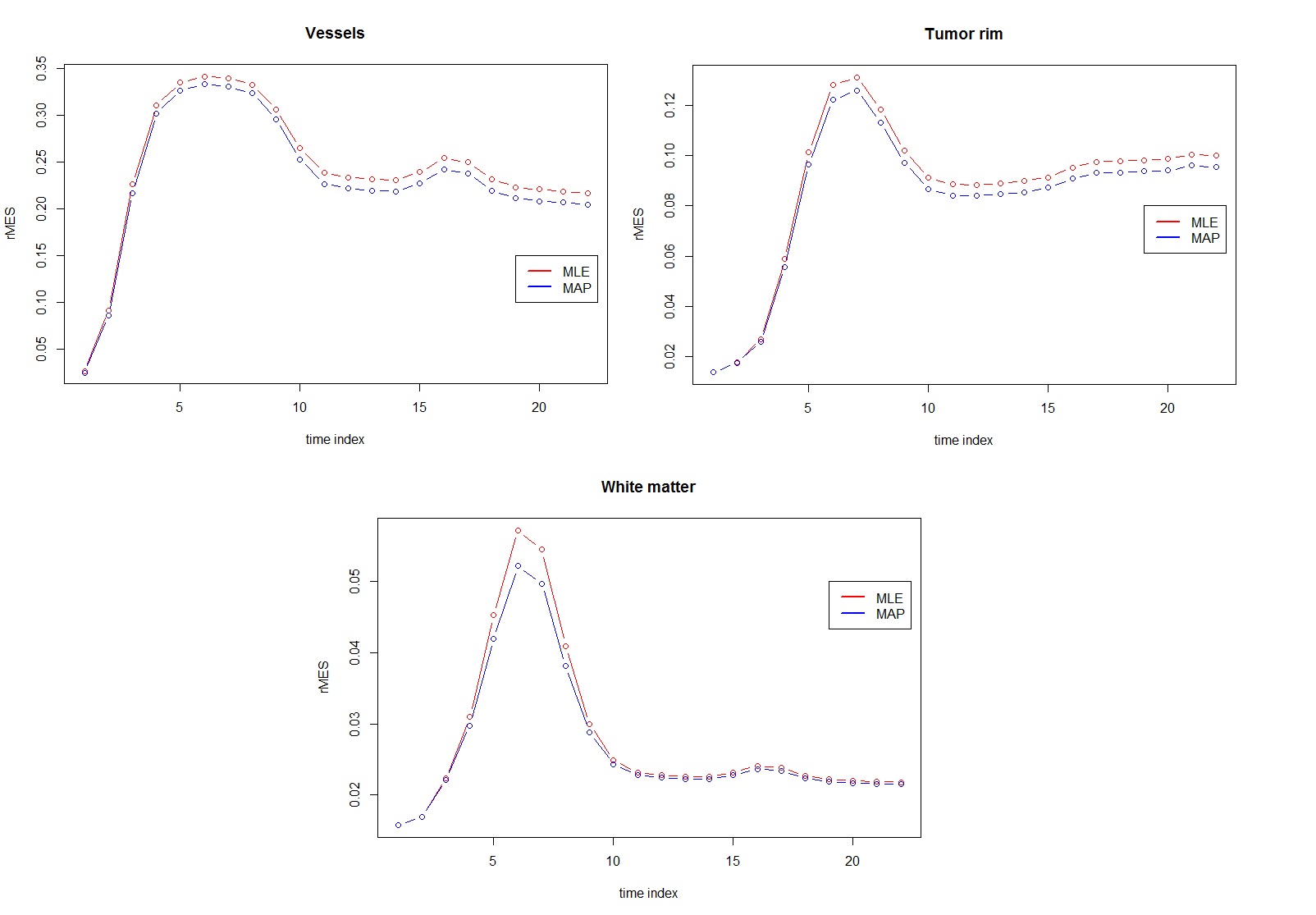}
    \caption{\small {rMSE comparison between MAP estimators and MLEs for vessels, tumor rim and white matter at 3T with rSNR=1 over 50 simulations}}
    \label{fig:5}
\end{figure}
Figure 4 is rMSE comparison between MAP estimators and MLEs with rSNR=1 at 3T and it also shows improvements compared to MLEs. Note that in general the rMSE in Figure 4 is larger than the
one in Figure 3 due to
 the different noise levels.
\section{Conclusion and Discussion}

From the analysis above, it shows improvements by using spatial information in vessels for smaller images. And the Leroux model outperforms the Besag models. The MAP estimators are better
than MLEs for vessels, tumor rim and white matter in terms of rMSE over 50 simulations for larger images with the assumption that spatial dependence is invariant over different image sizes,
even though $\hat\tau$s are not optimal.

Although the smaller images are not practical in clinic, the results show clear evidences that borrowing strength from neighbors can improve the accuracy of CA concentration. The further
analysis could be done in future by writing one's own codes to implement BHMs for larger images instead of using R-INLA. In this case, with $(1)$ and $(2^*)$ as the data model, Leroux model
as the process model, one can check if the spatial dependence is invariant or not and how $\tau$'s change over different image sizes. Also different magnetic strength, e.g.1.5T, and more
rSNRs can be analysed as in \citet{Brynolfsson2014}.

The other restriction of this study is that time dependence is not considered which results in a relatively simple statistical model and fast computational time. In reality, it is very
reasonable
 to incorporate time dependence into the BHMs, which is called spatio-temporal BHMs \citep{Cressie2011}. Figure 5 illustrates the spatio-temporal idea in a much concise way.
\begin{figure}
    \centering
    \includegraphics[width=0.8\linewidth]{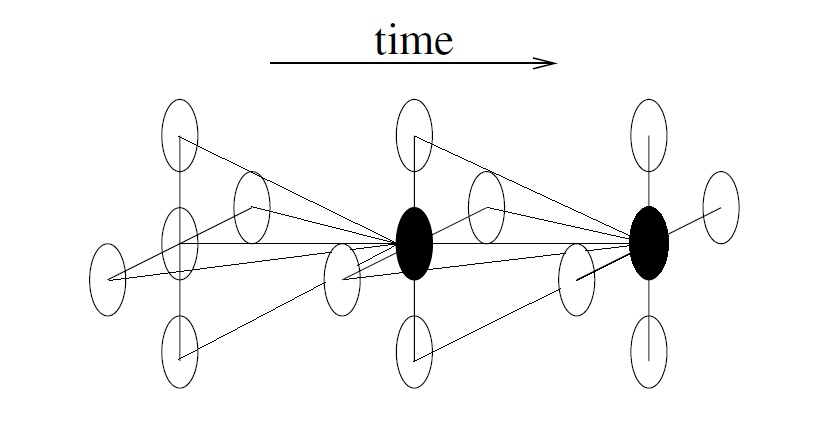}
    \caption{\small {Neighborhood structure in spatio-temporal model}}
    \label{fig:6}
\end{figure}
From the figure, the black dot is not only affected by its neighbors in space domain, but also by its neighbors in time domain. Many time series models can be used here, e.g random work
models and autoregressive models. By incorporating the time dependence, besides temporal trend, the interaction between spatial and temporal effects can be studied.

\section*{Acknowledgements}
This work was supported by the Swedish Research Council grant [Reg.No. 340-2013-5342]

\bibliographystyle{spbasic}      
\bibliography{ref}   

\end{document}